\pdfoutput=1

\documentclass[11pt]{article}

\usepackage[final]{acl}

\usepackage{times}
\usepackage{latexsym}

\usepackage[T1]{fontenc}

\usepackage[utf8]{inputenc}

\usepackage{microtype}

\usepackage{inconsolata}

\usepackage{graphicx}

\usepackage[svgnames]{xcolor}
\usepackage{amsmath,amsthm,amssymb}
\usepackage{mathrsfs}
\usepackage{tabularx}
\usepackage{multirow}
\usepackage{booktabs}
\usepackage{float}
\usepackage{makecell}
\usepackage{colortbl}
\usepackage{algorithm,algorithmic}
\usepackage{subfigure}

\title{Investigating Multi-layer Representations for Dense Passage Retrieval}

\author{Zhongbin Xie$^{1}$, Thomas Lukasiewicz$^{2,1}$ \\
         $^{1}$\,University of Oxford, UK \ $^{2}$\,Vienna University of Technology, Austria \\
         \texttt{zhongbin.xie@cs.ox.ac.uk}, \texttt{thomas.lukasiewicz@tuwien.ac.at}
}

\begin{document}
\maketitle
\begin{abstract}
Dense retrieval models usually adopt vectors from the last hidden layer of the document encoder to represent a document, which is in contrast to the fact that representations in different layers of a pre-trained language model usually contain different kinds of linguistic knowledge, and behave differently during fine-tuning. Therefore, we propose to investigate utilizing representations from multiple encoder layers to make up the representation of a document, which we denote Multi-layer Representations (MLR). We first investigate how representations in different layers affect MLR's performance under the multi-vector retrieval setting, and then propose to leverage pooling strategies to reduce multi-vector models to single-vector ones to improve retrieval efficiency. 
Experiments demonstrate the effectiveness of MLR over dual encoder, ME-BERT and ColBERT in the single-vector retrieval setting, as well as demonstrate that it works well with other advanced training techniques such as retrieval-oriented pre-training and hard negative mining.
\end{abstract}

\section{Introduction}

Dense passage retrieval is adopted in open-domain question answering~\citep{lee-etal-2019-latent,karpukhin-etal-2020-dense} to retrieve relevant passages from a large corpus for the reader model to extract answers. The dense retrieval technique encodes queries and documents into dense embeddings, and has wide applications in other knowledge-intensive tasks~\citep{petroni-etal-2021-kilt} as well as retrieval-augmented generation~(RAG, \citealp{lewis-2020-neurips,zhao-rag-2024}). 
Dense retrieval enjoys many advantages over sparse retrieval methods~\citep{tfidf,bm25}, such as alleviation of the term mismatch problem~\citep{furnas-1987-term}, and improved retrieval performance through supervised learning~\citep{karpukhin-etal-2020-dense}.

\begin{figure*}[htbp]
    \centering
    \begin{minipage}{0.499\linewidth}
        \centering
        \includegraphics[width=0.99\linewidth]{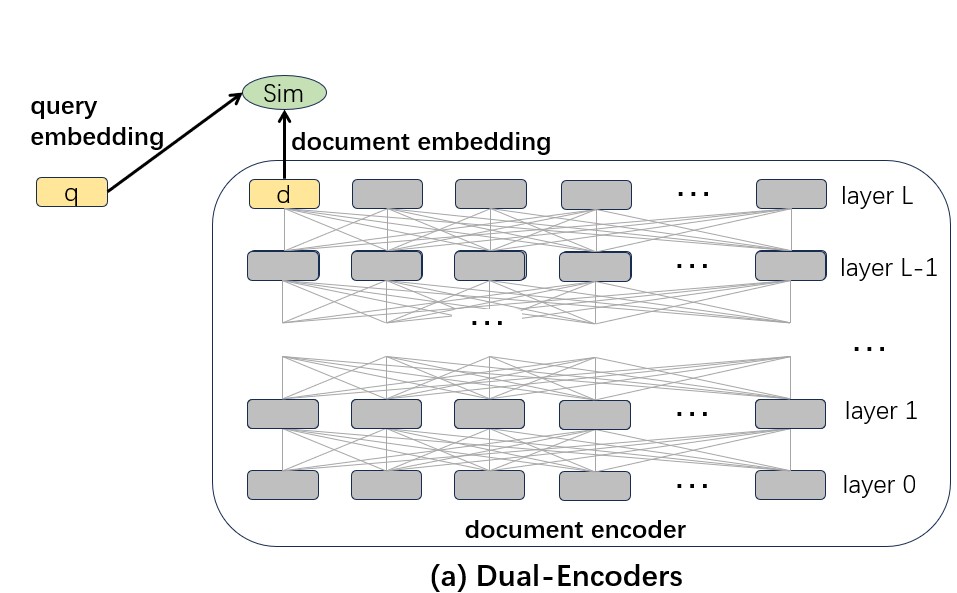}
    \end{minipage}
    \begin{minipage}{0.49\linewidth}
        \centering
        \includegraphics[width=0.99\linewidth]{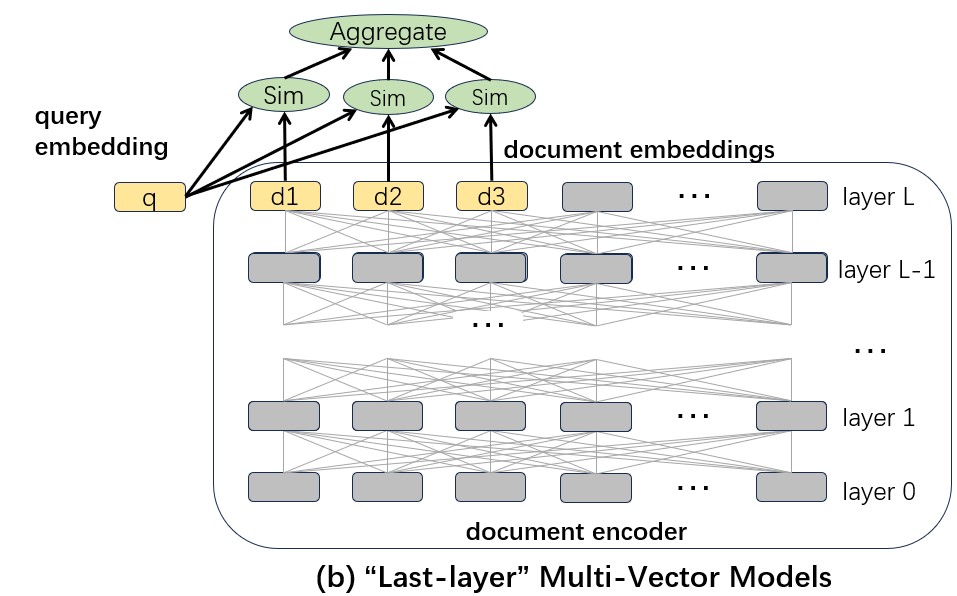}
    \end{minipage}
    \begin{minipage}{0.49\linewidth}
        \centering
        \includegraphics[width=0.99\linewidth]{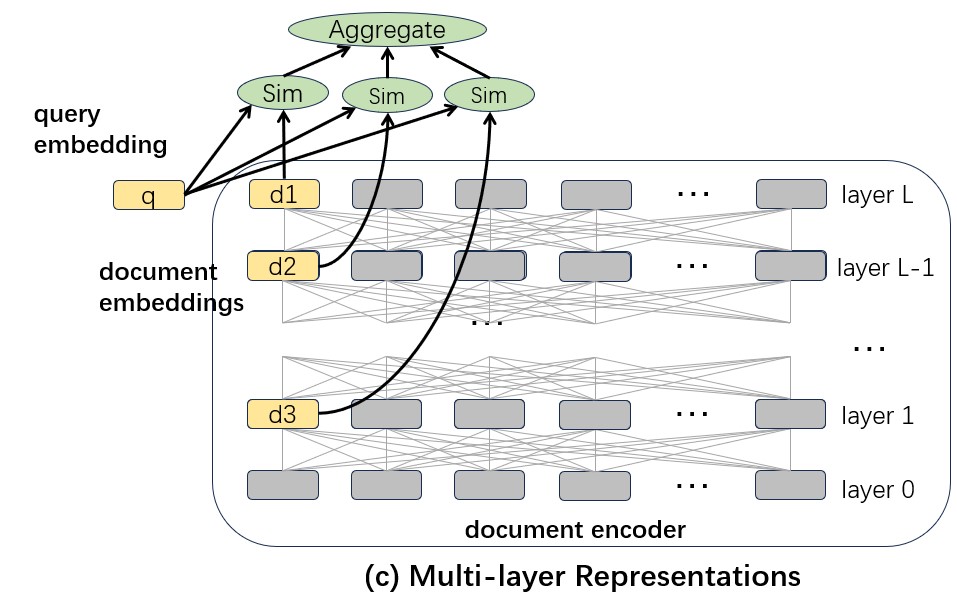}
    \end{minipage}
    \begin{minipage}{0.49\linewidth}
        \centering
        \includegraphics[width=0.99\linewidth]{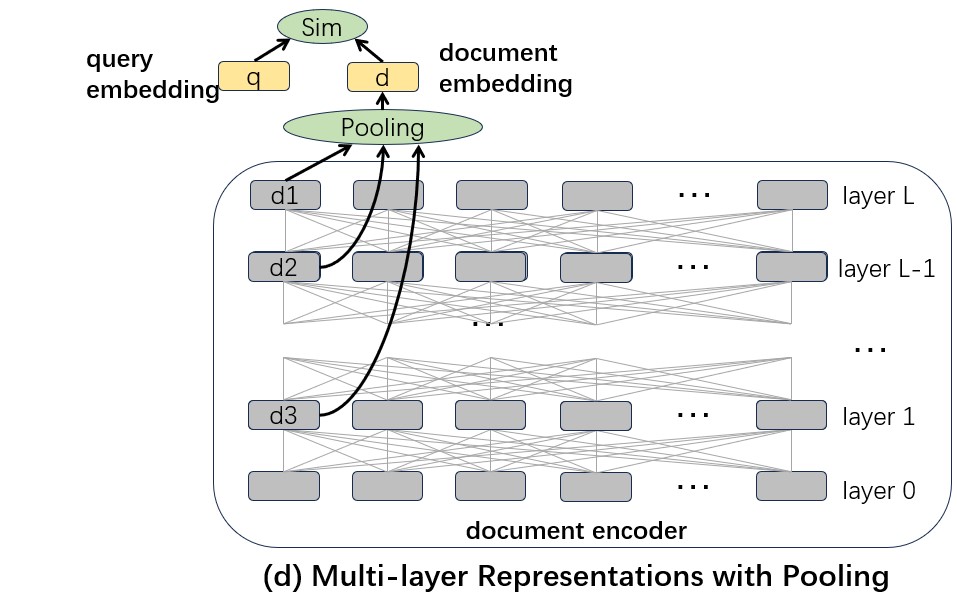}
    \end{minipage}
    \caption{Illustrations of different dense retrieval models. Yellow boxes represent query/document embedding vectors; green ellipses represent operations. Details of (c) and (d) are described in \S~\ref{sec:model}.}
    \label{fig:arch}
\end{figure*}

Dense retrieval models typically adopt a dual-encoder 
architecture (\citealp{karpukhin-etal-2020-dense}, cf.\ Figure~\ref{fig:arch} (a)), where the query and document are encoded by two encoders usually fine-tuned from a pre-trained language model. However, current dense retrieval architectures~\citep{colbert-2020,zhang-etal-2022-multi,hong-etal-2022-sentence} represent documents with vectors\footnote{We use the term ``embedding'' and ``vector'' interchangeably in this paper.} taken only from the encoder's last hidden layer (Figure~\ref{fig:arch} (b)). This is in contrast to studies which have shown that, for a pre-trained language model like BERT~\citep{devlin-etal-2019-bert}, representations in different layers contain different kinds of linguistic knowledge, and behave differently during fine-tuning~\citep{rogers-etal-2020-primer}. For example, syntactic information resides mainly in the middle layers of BERT, while the semantics spreads across all the layers~\citep{hewitt-manning-2019-structural,jawahar-etal-2019-bert,tenney-etal-2019-bert}, and the final layers are most task-specific after fine-tuning~\citep{liu-etal-2019-linguistic,kovaleva-etal-2019-revealing,hao-etal-2019-visualizing}. Given this observation, we propose to investigate utilizing representations from multiple encoder layers, instead of those only from the last layer, to make up the representation of a document, which we denote Multi-layer Representations (MLR, Figure~\ref{fig:arch} (c)). 

A straightforward way to utilize representations from multiple encoder layers to represent a document is to retrieve in a multi-vector setting, which is illustrated in Figure~\ref{fig:arch} (c). Unlike the vanilla dual-encoders where each document is represented by a \textit{single} vector, multi-vector models~\citep{colbert-2020,luan-etal-2021-sparse,zhang-etal-2022-multi,hong-etal-2022-sentence} represent each document with \textit{multiple} vectors, and thus enjoy more representational capacity. We propose to investigate how representations in different layers affect this representational capacity, and how this compares to previous ``last-layer'' multi-vector models like ME-BERT~\citep{luan-etal-2021-sparse}.

On the other hand, multi-vector models' improved retrieval performance over single-vector models usually comes with the cost of a decreased retrieval efficiency, because the number of document embeddings to be searched is proportional to the average number of vectors used to represent each document.
For example, for an 8-vector ME-BERT~\citep{luan-etal-2021-sparse}, it takes 481GB disk space to store the document embeddings for $\sim$21M documents, as well as 3.0321 seconds per query to build and retrieve the entire index. In contrast, for a single-vector dual-encoder, it only takes 60GB and 0.2056 seconds on the same machine. Given this, we further propose to pool the multi-vector representation of a document into a single one so that multi-vector models can be reduced to single-vector models during inference time (Figure~\ref{fig:arch} (d)). Thus, we can enjoy an improved retrieval performance over a single-vector dual-encoder with exactly the same retrieval efficiency.

Our contributions are as follows:
\begin{itemize}
    \item We propose to utilize representations from multiple encoder layers to represent a document, which is denoted as Multi-layer Representations (MLR). We investigate how representations in different layers affect MLR's performance under the multi-vector retrieval setting, and find that, with the last few layers and a relatively small vector number, MLR can effectively outperforms baselines with both BERT and T5. But unlike ME-BERT, MLR's representational capacity cannot scale up well with more representation vectors.
    
    \item We further propose to leverage pooling strategies to reduce multi-vector models to single-vector ones to improve retrieval efficiency. We explore self-contrastive pooling, average pooling, and scalar mix pooling, and find that single-vector MLR can outperform a single-vector dual encoder by a large margin.
    
    \item We demonstrate on diverse in-domain and out-of-domain retrieval datasets that single-vector MLR works well with other advanced training techniques such as retrieval-oriented pre-training and hard negative mining.\footnote{The code of this paper is available at \url{https://github.com/x-zb/mlr}.}
\end{itemize}

\section{Related Work}

\textbf{Single-vector retrieval models.} DPR~\citep{karpukhin-etal-2020-dense} adopts a dual-encoder architecture and a contrastive loss to learn dense representations for queries and documents. Improved training techniques are further developed to learn better single-vector representations, and can be roughly divided into three categories: (\romannumeral1) hard negative mining~\citep{xiong-2021-iclr,sun-etal-2022-reduce}, (\romannumeral2) retrieval-oriented pre-training~\citep{gao-callan-2021-condenser,gao-callan-2022-unsupervised,xiao-etal-2022-retromae,liu-etal-2023-retromae}, and (\romannumeral3) knowledge distillation from cross encoders~\citep{hofsttter-2020-distillation,qu-etal-2021-rocketqa,ren-etal-2021-rocketqav2,tao-etal-2024-adam}. These techniques are in general orthogonal to our proposed method, and we empirically investigate MLR's compatibility with retrieval-oriented pre-training and hard negative mining in \S~\ref{sec:integration}. 
 
\paragraph{Multi-vector retrieval models.}

Dual-encoders represent each document as a single vector, thus have limited representational power for long documents~\citep{luan-etal-2021-sparse}, are prone to be overfitting~\citep{menon-2022-icml}, and struggle to handle the one-to-many scenario where one document contains answers to multiple different queries~\citep{zhang-etal-2022-multi,hong-etal-2022-sentence}. Therefore, several multi-vector models have been proposed. 
Specifically, ColBERT~\citep{colbert-2020,santhanam-etal-2022-colbertv2} adopts representations of all the tokens in a document, while ME-BERT~\citep{luan-etal-2021-sparse} and MVR~\citep{zhang-etal-2022-multi} adopt a fixed number of vectors which are much fewer than the document length. \citet{tang-etal-2021-improving} cluster the token representations and adopt the resulting cluster centers to represent the document. \citet{hong-etal-2022-sentence} segment a document into sentences and  for each sentence introduce a learnable token, whose representations are then used to represent the document. All these models use representations from the document encoder's last layer, while we propose to explore the performance of intermediate layers.

\paragraph{Utilizing intermediate layers in deep learning.} There are also related work on utilizing multiple hidden layers of a neural network to make predictions~\citep{yan-2015-iccv,huang-2018-iclr,wehrmann-2018-icml,manginas-etal-2020-layer,evci-2022-icml}. For example, 
\citet{manginas-etal-2020-layer} leverage different layers of BERT representations for hierarchical multi-label document classification, while \citet{hosseini-etal-2023-bert} investigate BERT layers combination for semantic textual similarity. 
In information retrieval, \citet{nie-2018-sigir} aggregate the matching score of a query-document pair from different layers of a convolutional neural network, but their model focuses on the reranking task, and cannot be applied to large-scale retrieval. \citet{ennen-etal-2023-hierarchical} leverage hierarchical representations of a BERT encoder to represent a query (not a document as we do), and require to dynamically adjust the document index during search. 
Moreover, they report negative results on dense passage retrieval.
In contrast, we directly leverage multi-layer representations from a pre-trained language model to represent a document, and demonstrate its effectiveness in different scenarios.

\section{Multi-layer Representations}
\label{sec:model}

Given a text query $q$, we aim to find the relevant documents from a large document collection $\mathcal{D}=\{d_1, d_2, \ldots, d_N\}$, where $N$ can range from millions to billions. We adopt one query encoder $E_Q(\cdot)$ and one document encoder $E_D(\cdot)$ to represent the corresponding text sequences as real-valued dense vectors.. 
For the query/document encoders, we will experiment with two pre-trained language models, BERT~\citep{devlin-etal-2019-bert} and T5~\citep{t5-paper}. For convenience, we will describe our method using BERT, and highlight the adaptations for T5 in \S~\ref{sec:t5}.

For a query $q$, the \textit{[CLS]} representation in the last layer is adopted as the query embedding $E_Q(q) \doteq h_q\in\mathbb{R}^D$. For a document $d$ with $T$ tokens, assume the output of the document encoder with $L$ transformer layers is a set of hidden states $\{h_i^{(l)}\in\mathbb{R}^D\ |\ l=0,1,2,\dots,L;i=0,1,2,\ldots,T\}$, where $h_i^{(l)}$ denotes the hidden state in layer $l$ at position $i$ (layer 0 denotes the word embedding layer, and position 0 denotes the \textit{[CLS]} token). For a dual-encoder, $E_D(d) \doteq h_0^{(L)}$; for ME-BERT~\citep{luan-etal-2021-sparse}, $E_D(d) \doteq \{h_i^{(L)}\ |\ i=0,1,\dots,m-1\}$, where $m$ is the number of representation vectors for each document; for ColBERT~\citep{colbert-2020}, $E_D(d) \doteq \{h_i^{(L)}\ |\ i=0,1,\dots,T\}$.

\subsection{Multi-vector retrieval}

BERT's representations in different layers contain different kinds of knowledge and behave differently during fine-tuning. Thus, to enrich our document representation, we leverage the \textit{[CLS]} representations $h_0^{(l)}$ from different layers to represent a document, as shown in Figure~\ref{fig:arch} (c). Specifically, we adopt
\begin{equation}
    E_D(d) \doteq \{h_0^{(l)}\ |\ l\in\mathcal{S}=\{l_1,l_2,\dots,l_m\}\}
    \label{equ:set}
\end{equation}
as our multi-vector representation for a document $d$. Here,  $\mathcal{S}=\{l_1,l_2,\dots,l_m\}\subseteq\{1,2,\dots,L\}$, and we always include the last layer (i.e., $l_m=L$) to leverage the full depth of the encoder.

We define the similarity between a query and a document as the maximum inner product between the query embedding $h_q$ and all of the document embeddings in $E_D(d)$:
\begin{equation}
    \text{sim}(q,d) = \max_{h_0^{(l)}\in E_D(d)}{h_q^{\mathsf{T}}h_0^{(l)}}.
    \label{equ:sim}
\end{equation}
The number of representation vectors per document (i.e., $m$), as well as how to choose the specific layers for a fixed $m$, are the keys for model performance. In \S~\ref{sec:multi-vector}, we will investigate different strategies including using last few layers, first few layers, and uniformly distributed layers. 

\paragraph{Training.} Given a training instance $\langle q,d^+,d^-_1,$ $\dots,d^-_n \rangle$, where $d^+$ is a positive (relevant) document, and the $d^-_i$s are $n$ negative (irrelevant) documents, we adopt the following contrastive loss to optimize encoder parameters:
\begin{align}
    & \mathcal{L}( q,d^+,d^-_1,\dots,d^-_n) \notag \\
    & = -\log{\frac{e^{\text{sim}(q,d^+)}}{e^{\text{sim}(q,d^+)}+\sum_{i=1}^n{e^{\text{sim}(q,d^-_i)}}}}\,.
    \label{equ:contrastive-loss}
\end{align}

In practice, we follow \citet{karpukhin-etal-2020-dense} to include one negative passage for each query, and adopt the in-batch negatives technique, where all the (positive and negative) documents corresponding to other queries in the same mini-batch are used as negative documents for this query. Thus, the number of negatives $n$ in Eq.~(\ref{equ:contrastive-loss}) equals to $2(B-1)+1$ with $B$ being the batch size.

\paragraph{Inference.} We adopt the \texttt{FAISS} library~\citep{johnson-2019-faiss} to index all the document vectors and conduct nearest neighbor search. Unlike models like ColBERT~\citep{colbert-2020} and DCSR~\citep{hong-etal-2022-sentence} where the number of representation vectors is different for each document, our model adopts a fixed number of vectors $m$. So,  when retrieving top-$k$ documents, we can simply retrieve top-$km$ vectors, and map them back to document ids by dividing each vector id with $m$ and taking the integer part. Finally, since we adopt the maximum dot product in Eq.~(\ref{equ:sim}), we can just merge the same document ids and take the top-$k$ results.

 \begin{figure*}[ht]
    \centering
    \begin{minipage}{0.499\linewidth}
        \centering
        \includegraphics[width=0.99\linewidth]{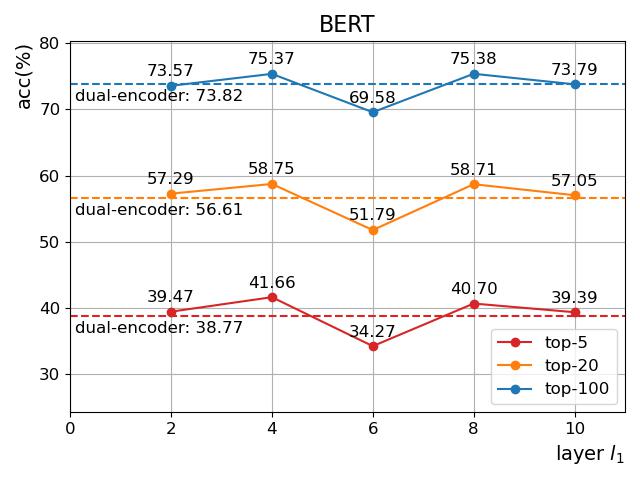}
    \end{minipage}
    \begin{minipage}{0.49\linewidth}
        \centering
        \includegraphics[width=0.99\linewidth]{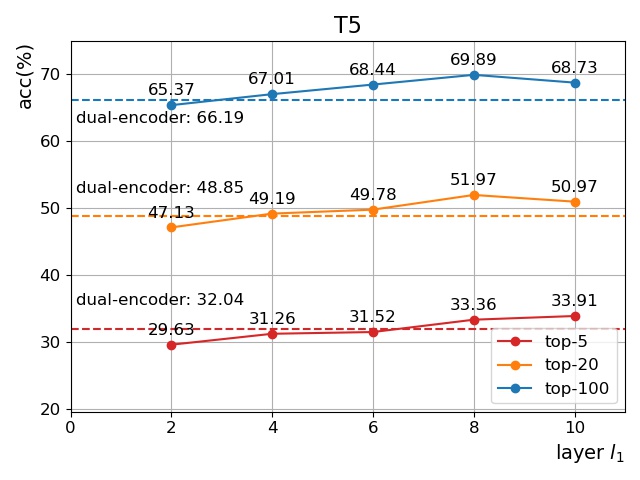}
    \end{minipage}
    \caption{Multi-vector retrieval accuracy w.r.t. different choices of $l_1$ in a 2-vector MLR model ($\mathcal{S}=\{l_1,l_2=12\}$). Results are evaluated on the SQuAD test set with BERT or T5.
    }
    \label{fig:dmde2}
\end{figure*}

\begin{table*}[h]
    \centering
    \footnotesize
    \begin{tabular}{l|ccc|ccc}
    \toprule
    \multirow{2}*{}&\multicolumn{3}{c|}{BERT}&\multicolumn{3}{c}{T5} \\
    {Layer Combinations} & {Top-5} & {Top-20} & {Top-100} & {Top-5} & {Top-20} & {Top-100} \\
    \hline
    {$\mathcal{S}_\text{last}=\{9,10,11,12\}$} & {41.63} & {58.68} & {74.72} & {31.99} & {49.51} & {68.76} \\
    {$\mathcal{S}_\text{first}=\{1,2,3,12\}$} & {38.54} & {55.90} & {72.87} & {27.28} & {44.30} & {63.60} \\
    {$\mathcal{S}_\text{uniform}=\{3,6,9,12\}$} & {34.62} & {52.39} & {70.11} & {32.88} & {51.04} & {69.15} \\
    \bottomrule
    \end{tabular}
    \caption{Impact of different layer combinations evaluated with a 4-vector MLR model on the SQuAD test set.}
    \label{tab:layer-dist}
\end{table*}

\subsection{Reducing to single-vector models through pooling}

\label{sec:pooling}

We further investigate if pooling multiple vectors to a single vector can achieve an improved performance in the single-vector retrieval setting. This is desirable, since single-vector retrieval is much more efficient than multi-vector retrieval, in terms of both space and time complexity, because the number of document embeddings to be searched is $mN$,  which is proportional to the number of vectors used to represent each document. 

Recall that $E_D(d)=\{h_0^{(l)}\ |\ l\in\mathcal{S}\}$ is a set of $m$ vectors that we use to represent a document, and assume that $h_D(d)$ is the single vector pooled from $E_D(d)$ to represent a document during inference. We propose the following \textbf{self-contrastive pooling} strategy: 

During inference, we simply take the last layer \textit{[CLS]} representation $h_0^{(L)}$ as the single vector representation for a document:
\begin{equation}
    h_D(d)=h_0^{(L)}.
\end{equation}
During training, we adopt the same maximum inner product similarity as Eq.~(\ref{equ:sim}) for negative documents, but use the inner product of $h_q$ and $h_D(d^+)$ for positive documents, resulting in the following contrastive loss:
\begin{align}
    & \mathcal{L}_{con}( q,d^+,d^-_1,\dots,d^-_n) \notag \\ 
    & = -\log{\frac{e^{h_q^{\mathsf{T}}h_D(d^+)}}{e^{h_q^{\mathsf{T}}h_D(d^+)}+\sum_{i=1}^n{e^{\text{sim}(q,d^-_i)}}}}\,.
    \label{equ:sc-loss}
\end{align}
Here, for positive documents, the vector whose inner product with $h_q$ is maximized is the same with that used during inference (i.e., $d_D(d^+)$). For negative documents, since we are minimizing the maximum inner product, the inner product for $h_D(d^-)$ is also minimized. Thus, Eq.~(\ref{equ:sc-loss}) can guarantee that our training and inference targets are the same.

To enhance $h_q^{\mathsf{T}}h_D(d^+)$ approaching $\text{sim}(q,d^+)$ such that Eq.~(\ref{equ:sc-loss}) is closer to Eq.~(\ref{equ:contrastive-loss}), we also add the following self-contrastive loss for positive documents as a regularization term:
\begin{align}
    & \mathcal{L}_{reg}(q,d^+)=-\log{\frac{e^{h_q^{\mathsf{T}}h_D(d^+)}}{\sum \limits_{h\in E_D(d^+)}{e^{h_q^{\mathsf{T}}h}}}}.
    \label{equ:reg-loss}
\end{align}
Thus, our final loss function is
\begin{align}
    & \mathcal{L}_{con}( q,d^+,d^-_1,\dots,d^-_n)+\lambda\mathcal{L}_{reg}(q,d^+),
    \label{equ:total-loss}
\end{align}
where $\lambda$ is a hyperparameter controlling the strength of the regularization. 

Besides, we also experiment with the following two simple pooling strategies: (\romannumeral1) \textbf{average pooling}, where we take the average vector of all the vectors in $E_D(d)$, and use it during training and inference as a single vector model; and (\romannumeral2) \textbf{scalar mix pooling}, where we take the weighted average of all the vectors in $E_D(d)$:
\begin{equation}
    h_D(d)=\sum_{h_0^{(l)}\in E_D(d)}{\text{softmax}(\alpha)_l h_0^{(l)}},
\end{equation}
with $\alpha \in\mathbb{R}^L$ being a set of learnable parameters for each layer. Similar methods are used to aggregate representations in ELMo~\citep{peters-etal-2018-deep}.

\subsection{T5 encoders}
\label{sec:t5}
For the dual-encoder architecture, since T5 does not have a \textit{[CLS]} token whose representations can be used as a document embedding, we follow \citet{ni-etal-2022-large} to leverage the average of all the token representations in the last layer of a T5 encoder as the document embedding . For our Multi-layer Representations architecture, we adopt the average token vectors in each selected layer as the multi-vector representations of a document.

\section{Experiments}


\subsection{Experimental setup}

\paragraph{Datasets and metrics.} In our experiments, we adopt the Natural Questions~\citep{kwiatkowski-etal-2019-natural}, TriviaQA~\citep{joshi-etal-2017-triviaqa}, and SQuAD~\citep{rajpurkar-etal-2016-squad} datasets processed by \citet{karpukhin-etal-2020-dense}. In each of the three datasets, we train on the training questions each of which is attached with one positive and one negative passage sampled from a pool of $\sim$100 BM25 negatives; we use the dev set for validation and report the top-$5/20/100$ accuracy on the test set. Top-$k$ accuracy is defined as the fraction of questions whose positive passages appear in the top-$k$ retrieved passages by the model. The number of questions in each dataset is shown in Table~\ref{tab:dataset} in Appendix~\ref{app:training-details}. The document collection consists of 21,015,324 Wikipedia passages, which are disjoint text blocks of 100 words.

\begin{figure*}[ht]
    \centering
    \begin{minipage}{0.499\linewidth}
        \centering
        \includegraphics[width=0.99\linewidth]{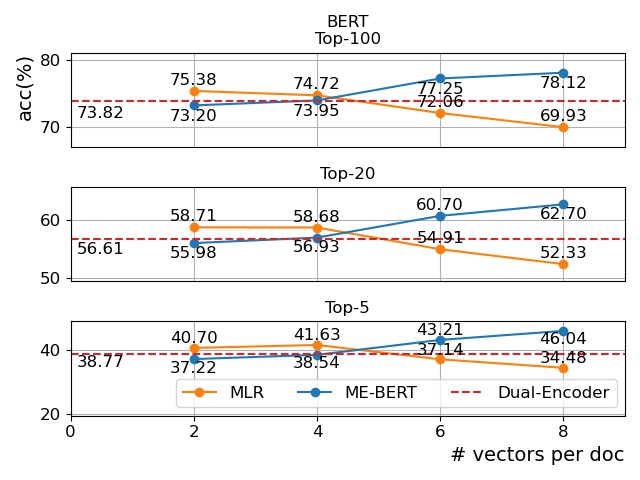}
    \end{minipage}
    \begin{minipage}{0.49\linewidth}
        \centering
        \includegraphics[width=0.99\linewidth]{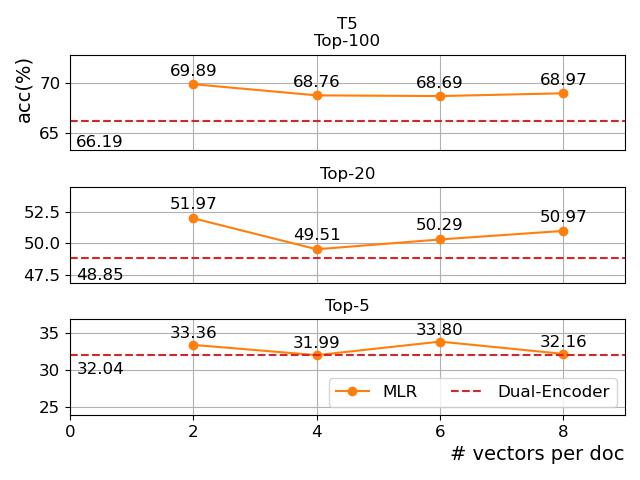}
    \end{minipage}
    \caption{Multi-vector retrieval accuracy w.r.t. the number of vectors per document (i.e., $m$). Results are evaluated on the SQuAD test set with BERT or T5.}
    \label{fig:dmde_me}
\end{figure*}

\paragraph{Training and implementation details.} We adopt \texttt{bert-base-uncased} or \texttt{google-t5/t5-base} from Huggingface~\citep{wolf-etal-2020-transformers} as our initial encoder, both of which consist of $L=12$ transformer layers. Training hyperparameters roughly follow those in \citet{karpukhin-etal-2020-dense}, where we use AdamW~\citep{adamw-paper} to train our models with an initial learning rate of 2e-5 for BERT (5e-4 for T5) and batch size of 128 for 40 epochs. In self-contrastive pooling, we search the regularization strength $\lambda$ in Eq.~(\ref{equ:reg-loss}) from $\{0.01,0.1,1,10\}$. Further details are in Appendix~\ref{app:training-details}. We adopt the gradient caching technique~\citep{gao-etal-2021-scaling} to use large batch sizes with restricted GPU memories. Our experiments are run on either four Quadro RTX 6000 or four Tesla V100 GPUs. During inference, we first divide the passage collection into 15 shards and encode each of them into a vector file; we then search an \texttt{IndexFlatIP} index built from one of the 15 shards so that it can be fed into the memory of four GPUs; we adopt a heap to keep the top results from each shard and merge them to get the final retrieval results.

\subsection{Multi-vector retrieval}
\label{sec:multi-vector}


\subsubsection{The impact of layer combinations}
\label{sec:layer-dist}

We first study a 2-vector MLR model where each document is represented by two vectors from layers $\mathcal{S}=\{l_1,l_2=12\}$ in Eq.~(\ref{equ:set}). Results with different choices of $l_1$ are shown in Figure~\ref{fig:dmde2}. We can see that, due to different network architectures and training objectives, BERT and T5 exhibit different performance patterns w.r.t. layer combinations. However, for both BERT and T5, the last few layers ($l_1=8, 10$) consistently outperform the dual-encoder baseline.  

Next, we focus on a 4-vector MLR model, and test three kinds of layer combinations: last four layers where $\mathcal{S}_\text{last}=\{9,10,11,12\}$, first three layers plus the last layer where $\mathcal{S}_\text{first}=\{1,2,3,12\}$, and uniformly distributed layers where $\mathcal{S}_\text{uniform}=\{3,6,9,12\}$. The results are shown in Table~\ref{tab:layer-dist}. We can see that the last few layers perform the best with BERT and comparable to uniformly distributed layers with T5.\footnote{We further test $m=6$ with T5, and confirm that the performance of last few layers (Top5/20/100 acc = 33.80/50.29/68.69) is more stable than that of uniformly distributed layers (Top5/20/100 acc = 26.71/44.03/63.03).} Therefore, we conclude that the last few layers perform better and more robustly in Multi-layer Representations.

\begin{table*}[h]
    \centering
    \footnotesize
    \begin{tabular}{p{0.277\textwidth}<{\raggedright}|p{0.041\textwidth}<{\centering}p{0.05\textwidth}<{\centering}p{0.058\textwidth}<{\centering}|p{0.041\textwidth}<{\centering}p{0.05\textwidth}<{\centering}p{0.058\textwidth}<{\centering}|p{0.041\textwidth}<{\centering}p{0.05\textwidth}<{\centering}p{0.058\textwidth}<{\centering}}
    \toprule
    \multirow{2}*{}&\multicolumn{3}{c|}{Natural Questions}&\multicolumn{3}{c|}{TriviaQA}&\multicolumn{3}{c}{SQuAD} \\
    {Models} & {\fontsize{7.8}{8}\selectfont Top-5} & {\fontsize{7.8}{8}\selectfont Top-20} & {\fontsize{7.8}{8}\selectfont Top-100} & {\fontsize{7.8}{8}\selectfont Top-5} & {\fontsize{7.8}{8}\selectfont Top-20} & {\fontsize{7.8}{8}\selectfont Top-100} & {\fontsize{7.8}{8}\selectfont Top-5} & {\fontsize{7.8}{8}\selectfont Top-20} & {\fontsize{7.8}{8}\selectfont Top-100} \\
    \hline
    {Dual-Encoder} & {68.06} & {79.34} & {85.90} & {71.12} & {79.63} & {84.96} & {38.77} & {56.61} & {73.82} \\
    \rowcolor{gray!20}\multicolumn{10}{l}{Multi-layer Representations (MLR)} \\
    {Self-Contrastive ($\mathcal{S}\!=\!\{1,\dots,12\}$)} & \underline{68.67} & \underline{79.56} & \underline{86.26} & \bf{71.43} & {79.55} & {85.03} & {41.00} & {58.08} & {74.47} \\
    {Self-Contrastive ($\mathcal{S}\!=\!\{10,12\}$)} & \bf{68.73} & \bf{80.00} & {86.20} & \underline{71.40} & {79.63} & \bf{85.28} & \bf{42.47} & \bf{59.93} & \bf{75.79} \\
    \rowcolor{gray!20}\multicolumn{10}{l}{ME-BERT} \\
    {Average Pooling} & {68.28} & {79.22} & \bf{86.40} & {71.12} & {79.28} & {84.96} & {41.33} & {58.94} & {75.36} \\
    {Scalar Mix Pooling} & {68.61} & {79.42} & {86.23} & {71.08} & {79.55} & {85.03}  & \underline{41.60} & \underline{59.01} & \underline{75.57} \\
    {Self-Contrastive} & {67.81} & {78.98} & {85.90} & {71.32} & \bf{79.70} & \underline{85.18} & {39.02} & {56.96} & {73.60} \\
    \rowcolor{gray!20}\multicolumn{10}{l}{ColBERT} \\
    {Average Pooling} & {68.03} & {79.22} & {85.82} & {71.02} & {79.40} & {85.05} & {39.74} & {57.37} & {73.82} \\
    {Self-Contrastive} & {68.03} & {79.36} & {86.18} & {71.31} & \underline{79.69} & {85.11} & {38.60} & {56.08} & {73.13} \\
    \bottomrule
    \end{tabular}
    \caption{Single-vector retrieval results on the Natural Questions, TriviaQA, and SQuAD's test sets. We compare the performance of applying the pooling strategies to different multi-vector models (i.e., MLR, ME-BERT, and ColBERT). Best and second best results are in \textbf{bold} and \underline{underlined}, respectively.}
    \label{tab:single-1}
\end{table*}

\subsubsection{The impact of vector numbers}

We next investigate MLR's performance with different number of representation vectors per document, i.e., $m$ in Eq.~(\ref{equ:set}). Specifically, we set $m=2,4,6,8$, and for each $m$, we adopt the best performing layer combination according to the analysis in \S~\ref{sec:layer-dist}, i.e., for $m=2$, we adopt $l_1=8$, and for $m=4,6,8$, we adopt the last few layers. We compare MLR with its ``last-layer'' counterpart, ME-BERT~\citep{luan-etal-2021-sparse}.\footnote{We also adapt ME-BERT for T5, but find that neither taking the first $m$ vectors nor taking the average vector plus the first $m-1$ vectors could outperform the dual-encoder baseline. Therefore we omit ME-BERT's performance with T5 in Figure~\ref{fig:dmde_me} and leave it for future work.} Both of the two multi-vector models adopt a fixed number of $m$ vectors to represent a document.

The results are shown in Figure~\ref{fig:dmde_me}. We can see that ME-BERT's performance increases with the increase of vector number $m$, which means that its representational capacity is increased as expected. For MLR, however, there is a decreasing trend in retrieval performance for both BERT and T5. This indicates that, unlike ME-BERT, MLR's representational capacity cannot scale up well with more representation vectors. This may be attributed to the fact that, Transformer representations in different layers at the same position are more correlated than those in the same layer but at different positions, and therefore adding too many vectors in MLR will limit vector diversity.

On the other hand, we notice that MLR can achieve superior performance over ME-BERT and dual-encoders with a relatively small number of vectors (e.g., $m=2$ or $4$) for both BERT and T5. We further examine which layer is selected on SQuAD's dev set for MLR with $m=2$ and $4$, and find that the last layer is always selected for all the queries. This means a last-layer query representation enhances a bias towards selecting the last-layer document representation, and the performance gains of MLR may largely come from the regularization effect of the max-aggregation during multi-vector training. This observation motivates us to further investigate whether pooling multiple vectors to a single vector through self-contrast could keep these performance gains.

\subsection{Single-vector retrieval}

In this section, we investigate the single-vector retrieval setting. 
Specifically, we adopt self-contrastive pooling for Multi-layer Representations, and experiment with pooling from all the layers ($\mathcal{S}\!=\!\{1,\dots,12\}$) and from $m=2$ layers ($\mathcal{S}\!=\!\{10,12\}$) in BERT.\footnote{Ablation studies on layer combinations and pooling strategies are in Appendix~\ref{app:ablation}.} 

\begin{table*}[ht]
    \centering
    \footnotesize
    \begin{tabular}{p{0.198\textwidth}<{\raggedright}|p{0.051\textwidth}<{\centering}p{0.06\textwidth}<{\centering}p{0.068\textwidth}<{\centering}|p{0.051\textwidth}<{\centering}p{0.06\textwidth}<{\centering}p{0.068\textwidth}<{\centering}|p{0.051\textwidth}<{\centering}p{0.06\textwidth}<{\centering}p{0.068\textwidth}<{\centering}}
    \toprule
    \multirow{2}*{}&\multicolumn{3}{c|}{Natural Questions}&\multicolumn{3}{c|}{TriviaQA}&\multicolumn{3}{c}{SQuAD} \\
    {} & {\fontsize{7.8}{8}\selectfont Top-5} & {\fontsize{7.8}{8}\selectfont Top-20} & {\fontsize{7.8}{8}\selectfont Top-100} & {\fontsize{7.8}{8}\selectfont Top-5} & {\fontsize{7.8}{8}\selectfont Top-20} & {\fontsize{7.8}{8}\selectfont Top-100} & {\fontsize{7.8}{8}\selectfont Top-5} & {\fontsize{7.8}{8}\selectfont Top-20} & {\fontsize{7.8}{8}\selectfont Top-100} \\
    \hline
    {BM25} & {--} & {59.1} & {73.7} & {--} & {66.9} & {76.7} & {--} & {68.8} & {80.0} \\
    {GAR} & {60.9} & {74.4} & {85.3} & {73.1} & {80.4} & {85.7} & {--} & {--} & {--} \\
    \hline
    {DPR$^\dagger$} & {68.1} & {79.3} & {85.9} & {71.1} & {79.6} & {85.0} & {38.8} & {56.6} & {73.8} \\
    {ANCE} & {--} & {81.9} & {87.5} & {--} & {80.3} & {85.3} & {--} & {--} & {--} \\
    {RocketQA} & {74.0} & {82.7} & {88.5} & {--} & {--} & {--} & {--} & {--} & {--} \\
    {DPR-PAQ} & {74.5} & {83.7} & {88.6} & {--} & {--} & {--} & {--} & {--} & {--} \\
    {Condenser} & {--} & {83.2} & {88.4} & {--} & {81.9} & {86.2} & {--} & {--} & {--} \\
    {coCondenser} & {75.8} & {84.3} & {89.0} & {76.8} & {83.2} & {87.3} & {--} & {--} & {--} \\
    {RetroMAE$^\dagger$} & {75.5}{\scriptsize$\pm$0.3} & {84.9}{\scriptsize$\pm$0.2} & {89.6}{\scriptsize$\pm$0.1} & {77.3}{\scriptsize$\pm$0.1} & \textbf{83.7}{\scriptsize$\pm$0.1} & \textbf{87.6}{\scriptsize$\pm$0.0} & {63.0}{\scriptsize$\pm$0.5} & {77.2}{\scriptsize$\pm$0.5} & {86.8}{\scriptsize$\pm$0.2} \\
    \hline
    {single-vector MLR} & \textbf{76.1}{\scriptsize$\pm$0.4}$^{*}$ & \textbf{85.0}{\scriptsize$\pm$0.1} & \textbf{89.7}{\scriptsize$\pm$0.1} & \textbf{77.4}{\scriptsize$\pm$0.1}$^{*}$ & \textbf{83.7}{\scriptsize$\pm$0.1} & \textbf{87.6}{\scriptsize$\pm$0.1} & \textbf{64.0}{\scriptsize$\pm$0.5}$^{*}$ & \textbf{77.7}{\scriptsize$\pm$0.4} & \textbf{86.9}{\scriptsize$\pm$0.1} \\
    \bottomrule
    \end{tabular}
    \caption{Single-vector retrieval results on the Natural Questions, TriviaQA, and SQuAD test sets for various baselines and models trained with retrieval-oriented pre-training and hard negative mining. $\dagger$: results are reproduced. For RetroMAE and single-vector MLR, we report the mean and standard deviation 
    from five runs with different random seeds. $*$ indicates the improvements of single-vector MLR over RetroMAE is statistically significant ($p<0.05$). Best results are in \textbf{bold}, and ``--'' means the results are unavailable in the original paper.}
    \label{tab:single-2}
\end{table*}

\subsubsection{Retrieval results}

Single-vector retrieval results of Multi-layer Representations ($\mathcal{S}\!=\!\{1,\dots,12\}$ \& $\mathcal{S}\!=\!\{10,12\}$), as well as those of ME-BERT~\citep{luan-etal-2021-sparse} and ColBERT~\citep{colbert-2020}\footnote{Different from the original ColBERT, here we only adopt the \textit{[CLS]} embedding for the query to avoid searching multiple times. This is to simplify the training and inference process for efficiency and consistency with other baselines.}, are shown in Table~\ref{tab:single-1}. For single-vector ME-BERT, we adopt $m=8$, since this is the best performing setting for multi-vector ME-BERT in  Figure~\ref{fig:dmde_me}; for single-vector ColBERT, scalar mix pooling is not applicable, since ColBERT adopts a non-fixed number of vectors to represent each document. 

First, on all three datasets, MLR with self-contrastive pooling can usually improve the retrieval accuracy over the dual-encoder baseline by a large margin. For example, on top-5 accuracy, self-contrastive pooling with $\mathcal{S}\!=\!\{10,12\}$ leads to a $+3.70\%$ improvement on SQuAD, and a $+0.67\%$ improvement on Natural Questions, while self-contrastive pooling with $\mathcal{S}\!=\!\{1,\dots,12\}$ can lead to a $+0.31\%$ improvement on TriviaQA. Notably, these improvements are made with the same inference time and space complexity as a dual-encoder model.

Second, compared to ME-BERT and ColBERT, MLR can achieve the best performance in most scenarios, which indicates that pooling representations from different layers is more effective than pooling those only in the last layer.

Note also that, compared to other training techniques like retrieval-oriented pre-training or hard negative mining, our single-vector MLR is simple and does not require additional complicated training stages. On the one hand, this means that when we restrict our total computation budget to only one training stage (this is usually desirable for training large-language-model-sized retrievers like repLLaMA~\citep{repllama-paper}), our method is still applicable, but the above two methods are not; on the other hand, when we have the budget to do more training stages, our method can be directly integrated with retrieval-oriented pre-training and hard negative mining, which is illustrated in \S~\ref{sec:integration} below.

\subsubsection{Integrating with retrieval-oriented pre-training and hard negative mining}
\label{sec:integration}

In this section, we integrate MLR in the single-vector retrieval setting with two advanced training techniques: retrieval-oriented pre-training and hard negative mining. Specifically, we adopt a MLR model with $\mathcal{S}\!=\!\{10,12\}$ and self-contrastive pooling. We follow \citet{gao-callan-2022-unsupervised}'s two-stage training procedure: in the first stage, the model is trained with BM25 negatives; then the trained model is used to mine hard negatives (i.e., top retrieved passages by the trained model in the first stage that do not contain the answer) 
for questions in the training set; in the second stage, the model is trained with the concatenation of the original and the mined training set. Models in both stages are initialized with a retrieval-oriented pre-trained checkpoint RetroMAE~\citep{xiao-etal-2022-retromae}. More training details are in Appendix~\ref{app:integration}.

For baselines, we compare to popular sparse (BM25 and GAR~\citep{mao-etal-2021-generation}) and dense (DPR~\citep{karpukhin-etal-2020-dense}, ANCE~\citep{xiong-2021-iclr}, RocketQA~\citep{qu-etal-2021-rocketqa}, DPR-PAQ~\citep{oguz-etal-2022-domain}, Condenser~\citep{gao-callan-2021-condenser} and coCondenser~\citep{gao-callan-2022-unsupervised}) retrieval systems.

From Table~\ref{tab:single-2}, we can see that our single-vector MLR can benefit the baseline dual encoders in most cases. For example, when compared to RetroMAE regarding top-5 accuracy, it leads to a $+0.6\%$ and $+1.0\%$ improvement on Natural Questions and SQuAD, respectively.

\begin{table}[ht]
    \centering
    \footnotesize
    \begin{tabular}{l|cc}
    \toprule
    \multirow{2}*{}&\multicolumn{2}{c}{MS MARCO Dev} \\
    {} & {\fontsize{7.8}{8}\selectfont MRR@10} & {\fontsize{7.8}{8}\selectfont Recall@1000} \\
    \hline
    {ANCE} & {33.0} & {95.9} \\
    {SEED} & {33.9} & {96.1} \\
    {coCondenser} & {38.2} & {98.4} \\
    {Aggretriver} & {36.3} & {97.3} \\
    {SPLADE-max} & {34.0} & {96.5} \\
    {SimLM (stage 2)} & {39.1} & {98.6} \\
    {RetroMAE (stage 2)} & {39.3} & {98.5} \\
    \hline
    {single-vector MLR (stage 1)} & {37.6} & {98.5} \\
    {single-vector MLR (stage 2)} & \bf{39.5} & \bf{98.7} \\
    \bottomrule
    \end{tabular}
    \caption{Single-vector retrieval results on the MS-MARCO dev set. Best results are in \textbf{bold}.
    }
    \label{tab:msmarco}
\end{table}

We further evaluate our method on the MS MARCO passage ranking dataset~\citep{msmarco-paper}, which contains 502,939 training queries. We report MRR@10 and Recall@1000 on its 6,980 dev queries (MS MARCO Dev), as well as NDCG@10 on the 43 test queries of the TREC 2019 Deep Learning Track (DL'19)~\citep{dl19-paper}. The size of the passage collection is 8,841,823. Although distillation from cross encoders can usually achieve superior performance~\citep{ren-etal-2021-rocketqav2,zhang-2022-iclr,tao-etal-2024-adam}, it is computationally much more expensive to train additional cross encoders and generate teacher scores for a large training set. Therefore, we focus on lightweight systems without knowledge distillation, and follow the same two-stage procedure in \S~\ref{sec:integration} to train MLR.
Training hyperparameters are the same as those in Appendix~\ref{app:training-details}, except that we adopt an initial learning rate of $1e-5$, total training epochs of 4, and 15 negatives per query for both stages 1 and 2. In stage 2, for each query, we take 100 negatives from the mined hard negatives to make up the negative pool. We adopt the last checkpoint for inference. $\lambda$ is set to $0.01$ for self-contrastive pooling.
For baselines, we compare to ANCE, SEED~\citep{lu-etal-2021-less}, coCondenser, Aggretriver~\citep{lin-etal-2023-aggretriever}, SPLADE-max~\citep{spladev2-paper}, SimLM (stage 2)~\citep{wang-etal-2023-simlm} and RetroMAE (stage 2)~\citep{xiao-etal-2022-retromae}. 

Results on the MS MARCO dev set are in Table~\ref{tab:msmarco}, where we can see that single-vector MLR outperforms RetroMAE by $0.2\%$ on both MRR@10 and Recall@1000. On DL'19 test set, single-vector MLR (stage 2) achieves an NDCG@10 of 68.8\% compared to RetroMAE's 69.9\%. 
We further find that, if we increase the regularization coefficient $\lambda$ from 0.01 to 1, single-vector MLR could achieve an NDCG@10 of 70.3\%, but with a decrease in MRR@10. Since the NDCG@10 metric emphasizes the first results that the users will see~\citep{dl19-paper}, we may adjust the regularization strength to fit different requirements. 

\subsubsection{Out-of-domain evaluation on BEIR}

We additionally conduct out-of-domain evaluation on the BEIR~\citep{beir-paper} benchmark. 
We initialize our model with RetroMAE's pre-trained checkpoint, and train the model on 502,939 MS MARCO~\citep{msmarco-paper} training queries, adopting an initial learning rate of $3e-5$, total training epochs of 10, and a $\lambda$ of $0.01$. 
In the BEIR benchmark, Signal 1M, Arguana, and Quora are considered to be symmetric tasks, where queries and documents are of about the same length and have the same amount of content. Since MLR is trained in an asymmetric manner (i.e., we encode documents and queries in different ways), here we evaluate on the other asymmetric datasets in BEIR.

Results are shown in Table~\ref{tab:beir}. We can see that MLR performs significantly better than the dual encoder-based RetroMAE on five datasets (by greater than 1\% in NDCG@10), while performs similarly with RetroMAE on the rest. This leads to a $+0.7\%$ improvements in average NDCG@10.

\begin{table}[h]
    \centering
    \footnotesize
    \begin{tabular}{l|cc}
    \toprule
    {} & {RetroMAE} & {single-vector MLR} \\
    \hline
    {NQ} & {0.508} & {0.515} \\
    {HotpotQA} & {0.627} & {0.620} \\
    {FiQA-2018} & {0.300} & {0.314} \\
    {TREC-NEWS} & {0.421} & {0.406} \\
    {Robust04} & {0.432} & {0.430} \\
    {Touche-2020} & {0.257} & {0.299} \\
    {CQADupStack} & {0.309} & {0.309} \\
    {DBPedia} & {0.390} & {0.390} \\
    {SCIDOCS} & {0.151} & {0.150} \\
    {FEVER} & {0.746} & {0.766} \\
    {Climate-FEVER} & {0.220} & {0.253} \\
    {SciFact} & {0.639} & {0.651} \\
    {TREC-COVID} & {0.767} & {0.764} \\
    {NFCorpus} & {0.307} & {0.305}  \\
    {BioASQ} & {0.414} & {0.406}  \\
    \hline
    {Average} & {0.432} & {0.439} \\
    \bottomrule
    \end{tabular}
    \caption{NDCG@10 results on BEIR for the dual encoder-based RetroMAE and our single-vector MLR.}
    \label{tab:beir}
\end{table}

\section{Conclusions}

In this study, we proposed to leverage representations from different encoder layers to represent a document in text retrieval.
For multi-vector retrieval,
we investigated how representations in different layers affect MLR's performance.
For single-vector retrieval, we found that MLR can outperform single-vector dual encoder by a large margin, and that pooling representations from different layers is more effective than pooling from representations only in the last layer. We also showed that advanced training techniques such as retrieval-oriented pre-training and hard negative mining can further boost MLR's performance.

\section*{Limitations}
 
Our current method is asymmetric in terms of representing queries and documents, i.e., we represent each document with multiple vectors and regularize them with the self-contrastive pooling loss, while only representing each query with a single vector. This is for efficiency consideration, but may limit our method's capacity and decrease its performance on symmetric search problems such as duplicate question identification and argument mining.
It would be a direct future work to explore utilizing multiple vectors to represent the query as well.    
Besides, the improved retrieval system may have broader societal impacts such as insufficient representations of minority groups, exhibiting stereotyped or biased results, and so on. Detection and mitigation of such unintended behaviors is of great significance but beyond the scope of this paper.

\section*{Acknowledgments}
This work was supported by the Alan Turing Institute under the EPSRC grant EP/N510129/1, by the AXA Research Fund, and by the EU TAILOR grant 952215.
We also acknowledge the use of Oxford’s ARC facility, of the EPSRC-funded Tier 2 facilities JADE (EP/P020275/1), and of GPU computing support by Scan Computers International Ltd.

\bibliography{custom,anthology}

\clearpage

\appendix

\section{Additional Training Details}
\label{app:training-details}

We adopt two separate BERT (or T5 encoder) from Huggingface~\citep{wolf-etal-2020-transformers} as our query and document encoder. Statistics of the adopted datasets are shown in Table~\ref{tab:dataset}. For all the experiments, we adopt a linear learning rate scheduler with warm up proportion $0.05$. We clip the gradient if it is larger than $2.0$. Maximum input sequence length for BERT (or T5 encoder) is set to 256. For the validation metric, we follow \citet{karpukhin-etal-2020-dense} to initially adopt cross entropy loss and then change to average rank on the dev set after 30 epochs. For Natural Questions and SQuAD, the best dev checkpoint is adopted for inference, while for TriviaQA, we adopt the last checkpoint. In self-contrastive pooling, we search the regularization strength $\lambda$ in Eq.~(\ref{equ:reg-loss}) from $\{0.01,0.1,1,10\}$, and adopt $\lambda=1$ for MLR ($\mathcal{S}=\{10,12\}$) and ME-BERT, $\lambda=0.1$ for MLR ($\mathcal{S}=\{1,\dots,12\}$), and $\lambda=0.01$ for ColBERT. It typically takes 16 to 30 hours to train one model on one dataset and encode the passage collection with four V100 GPUs, and 60 to 500GB disk space to store the encoded vector files. Due to long training time and large storage requirements, all the results in this paper (except those in \S~\ref{sec:integration}) are from a single run using seed 12345, which is consistent with \citet{karpukhin-etal-2020-dense} and most other works. We adhere to the licenses and intended use of the pre-trained checkpoints and datasets provided in their original papers or on their websites.

\begin{table}[b]
    \centering
    \footnotesize
    \begin{tabular}{lccc}
    \toprule
    \textbf{Dataset} & \textbf{Train (original / filtered)} & \textbf{Dev} & \textbf{Test} \\
    \hline
    {NQ} & {79,168 / 58,880} & {8,757} & {3,610} \\
    {TriviaQA} & {78,785 / 60,413} & {8,837} & {11,313} \\
    {SQuAD} & {78,713 / 70,096} & {8,886} & {10,570} \\
    \bottomrule
    \end{tabular}
    \caption{Number of questions in each dataset. For the training questions, we follow \citet{karpukhin-etal-2020-dense} to filter out questions with no associated positive passages, and the number of the remaining questions are presented after the slash. NQ stands for Natural Questions.}
    \label{tab:dataset}
\end{table}

\section{Additional Multi-vector Retrieval Results on Natural Questions}
\label{app:nq}

\begin{figure}[h]
    \centering
    \includegraphics[width=0.99\linewidth]{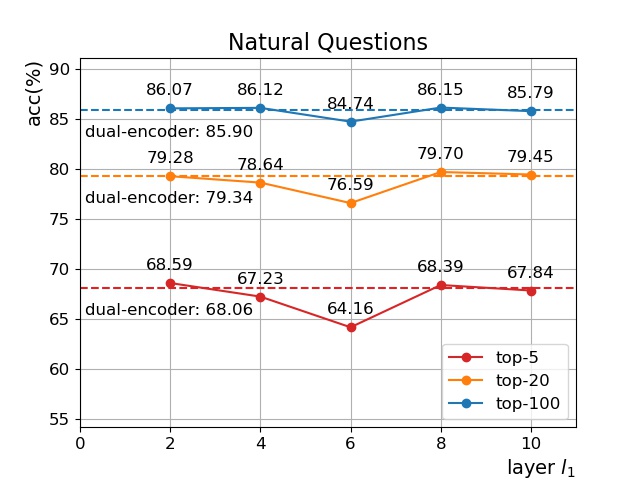}
    \caption{Multi-vector retrieval accuracy w.r.t. different choices of $l_1$ in a 2-vector MLR model ($\mathcal{S}=\{l_1,l_2=12\}$). Results are evaluated on the Natural Questions test set with BERT.}
    \label{fig:dmde2_nq}
\end{figure}

\begin{figure}[h]
    \centering
    \includegraphics[width=0.99\linewidth]{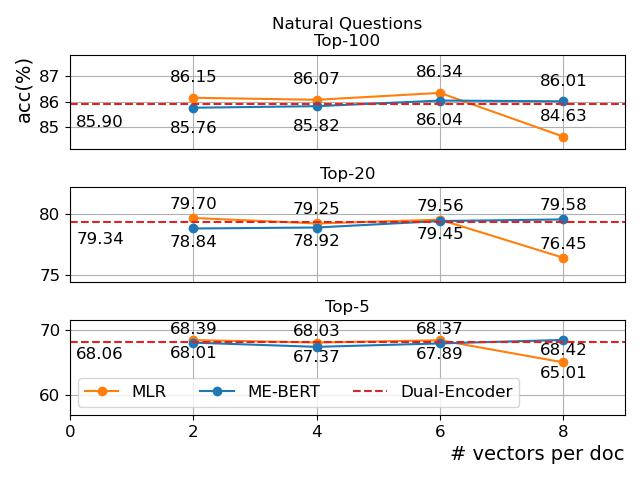}
    \caption{Multi-vector retrieval accuracy w.r.t. the number of vectors per document (i.e., $m$). Results are evaluated on Natural Questions test set with BERT.}
    \label{fig:dmde_me_nq}
\end{figure}

Additional multi-vector retrieval results on Natural Questions with BERT are provided in Figure~\ref{fig:dmde2_nq} and \ref{fig:dmde_me_nq}, which show similar trends to those on SQuAD with BERT (the left subgraph of Figure~\ref{fig:dmde2} and \ref{fig:dmde_me}).

\section{Ablation Studies on Pooling Strategies and Layer Combinations for Single-vector MLR}
\label{app:ablation}

\begin{table}[ht]
    \centering
    \footnotesize
    \begin{tabular}{l|ccc}
    \toprule
    \multirow{2}*{}&\multicolumn{3}{c}{Natural Questions} \\
    {Pooling Methods} & {Top-5} & {Top-20} & {Top-100} \\
    \hline
    \rowcolor{gray!20}\multicolumn{4}{l}{MLR ($\mathcal{S}\!=\!\{1,\dots,12\}$)} \\
    {Average} & {67.17} & {78.34} & {85.60} \\
    {Scalar Mix} & {66.81} & {78.59} & {85.43} \\
    {Self-Contrastive} & \bf{68.67} & \bf{79.56} & \bf{86.26} \\
    \rowcolor{gray!20}\multicolumn{4}{l}{MLR ($\mathcal{S}\!=\!\{10,12\}$)} \\
    {Average Pooling} & {68.03} & {79.42} & {85.93} \\
    {Scalar Mix Pooling} & {68.31} & {79.78} & {85.87} \\
    {Self-Contrastive} & \bf{68.73} & \bf{80.00} & \bf{86.20} \\
    \bottomrule
    \end{tabular}
    \caption{Single-vector retrieval accuracy w.r.t.\ different pooling methods for MLR with layer combinations $\mathcal{S}\!=\!\{1,\dots,12\}$ and $\mathcal{S}\!=\!\{10,12\}$. Best results for each layer combination are in \textbf{bold}.}
    \label{tab:pooling methods}
\end{table}

\begin{table}[ht]
    \centering
    \footnotesize
    \begin{tabular}{l|ccc}
    \toprule
    \multirow{2}*{}&\multicolumn{3}{c}{Natural Questions} \\
    {} & {Top-5} & {Top-20} & {Top-100} \\
    \hline
    {$\mathcal{S}\!=\!\{2,12\}$} & {67.98} & {79.47} & {85.96} \\
    {$\mathcal{S}\!=\!\{4,12\}$} & {67.78} & \underline{79.86} & {86.09} \\
    {$\mathcal{S}\!=\!\{6,12\}$} & \underline{68.34} & {79.50} & \bf{86.26} \\
    {$\mathcal{S}\!=\!\{8,12\}$} & {67.42} & {78.73} & {86.01} \\
    {$\mathcal{S}\!=\!\{10,12\}$} & \bf{68.73} & \bf{80.00} & \underline{86.20} \\
    \bottomrule
    \end{tabular}
    \caption{Single-vector retrieval accuracy w.r.t.\ different layer combinations for a 2-vector MLR model with self-contrastive pooling. Best and second best results are in \textbf{bold} and \underline{underlined}, respectively.}
    \label{tab:combinations}
\end{table}

We first compare different pooling strategies used in MLR. The results on Natural Questions are shown in Table~\ref{tab:pooling methods}, where we can see that self-contrastive pooling consistently performs better than average pooling and scalar mix pooling. 

Next, we study the performance of different layer combinations in the 2-vector MLR model with self-contrastive pooling. The results on Natural Questions are shown in Table~\ref{tab:combinations}, where we can see that pooling from layers $\mathcal{S}=\{10,12\}$ achieves the best top-$5$ and top-$20$ accuracy, while obtains a reasonable top-$100$ accuracy. Therefore we adopt $\mathcal{S}=\{10,12\}$ in our experiment.

\section{Training Details for Experiments with Retrieval-oriented Pre-training and Hard Negative Mining}
\label{app:integration}

Training hyperparameters are the same as those in Appendix~\ref{app:training-details}, except that, for stage 1, we adopt an initial learning rate of $5e-6$ and total training epochs of 40; for stage 2, an initial learning rate of $5e-6$ and total training epochs of 20. In stage 2, for each question, we take 50 negatives from each of the original and generated training set to make up a negative pool of 100 negatives. We adopt the last checkpoint for inference on Natural Questions and TriviaQA following ~\citet{gao-callan-2022-unsupervised}, and the best dev checkpoint for SQuAD. $\lambda$ is set to $0.1$ for self-contrastive pooling.

\end{document}